\documentclass[prl,aps,twocolumn,showpacs,superscriptaddress]{revtex4}
\usepackage{graphicx}
\usepackage{amsfonts}

\begin{document}

\title{Magnetic field induced lattice anomaly inside the superconducting state of CeCoIn$_5$: evidence of the proposed Fulde-Ferrell-Larkin-Ovchinnikov state}

\author{V. F. Correa}
\altaffiliation[Present address: ]{Comisi\'on Nacional de Energ\'{\i}a At\'omica, Centro At\'omico Bariloche, 8400 S. C. de Bariloche, Argentina}
\affiliation{National High Magnetic Field Laboratory, Florida State University, Tallahassee, Florida 32310, USA}

\author{T. P. Murphy}
\affiliation{National High Magnetic Field Laboratory, Florida State University, Tallahassee, Florida 32310, USA}

\author{C. Martin}
\affiliation{National High Magnetic Field Laboratory, Florida State University, Tallahassee, Florida 32310, USA}

\author{K. M. Purcell}
\affiliation{National High Magnetic Field Laboratory, Florida State University, Tallahassee, Florida 32310, USA}
\affiliation{Department of Physics, Florida State University, Tallahassee, Florida 32310, USA}

\author{E. C. Palm}
\affiliation{National High Magnetic Field Laboratory, Florida State University, Tallahassee, Florida 32310, USA}

\author{G. M. Schmiedeshoff}
\affiliation{Occidental College, Los Angeles, California 90041, USA}

\author{J. C.  Cooley}
\affiliation{Los Alamos National Laboratory, Los Alamos, New Mexico 87545, USA}

\author{S. W. Tozer}
\affiliation{National High Magnetic Field Laboratory, Florida State University, Tallahassee, Florida 32310, USA}

\date{\today}


\begin{abstract}

We report high magnetic field linear magnetostriction experiments on CeCoIn$_5$ single crystals. Two features are remarkable: 
(i) a sharp discontinuity in all the crystallographic axes associated with the upper superconducting critical field $B_{c2}$ that 
becomes less pronounced as the temperature increases; (ii) a distinctive second order-like feature observed only along the c-axis in the high field (10 T $ \lesssim B \leq  B_{c2}$) low temperature ($T \lesssim$ 0.35 K) region. 
This second order transition is observed only when the magnetic field lies within 20$^o$ of the ab-planes and there is no signature of it above $B_{c2}$, which raises questions regarding its interpretation as a field induced magnetically ordered phase.
Good agreement with previous results suggests that this anomaly is related to the transition to the Fulde-Ferrel-Larkin-Ovchinnikov superconducting state.

\end{abstract}

\maketitle

The unique properties of heavy fermion materials result from the strong correlation between the quasi-localized $f$-electrons and their interactions with the conduction $s,p$ and $d$-electrons. 
The partial delocalization resulting from this hybridization with the free electrons gives rise to large effective masses (100 - 1000 $m_e$) and a number of possible number of ground states. These include: unconventional magnetically mediated superconductivity, magnetic order, nonmagnetic Kondo singlet state, etc. Macroscopically, the magnitude of the correlations and eventually the ground state can be tuned using different parameters such as pressure or doping among others.

These features are beautifully exemplified in the CeMIn$_5$ (M = Ir, Rh, Co) family. Its tetragonal crystal structure alternates magnetic CeIn$_3$ and non-magnetic MIn$_2$ layers along the c-axis. Strong mass enhancement as well as reduced magnetic moments in the Ce ions due the Kondo effect are observed \cite{PagliusoB} at low temperatures. At ambient pressure, CeIrIn$_5$ is a superconductor ($T_c$ = 0.4 K) \cite{Petrovic1}, CeRhIn$_5$ is an antiferromagnet ($T_N$ = 3.8 K) \cite{Hegger} and, CeCoIn$_5$ is also a superconductor ($T_c$ = 2.3 K) \cite{Petrovic2}.
However,  these ground states are modified by external parameters. CeRhIn$_5$ shows pressure induced superconductivity ($T_c$ = 2.1 K at $P \sim$ 16 kbar) \cite{Hegger,Fisher} with a wide pressure range where both antiferromagnetism (AF) and superconductivity (SC) coexist \cite{Mito,Llobet,Nicklas,Knebel,Park}. Coexistence of AF and SC is also observed in CeRh$_{1-x}$Ir$_x$In$_5$ for 0.25 $ < x < $ 0.6 at ambient pressure \cite{Pagliuso1,Zheng} while two different superconducting phases can be detected under pressure \cite{Nicklas}. Coexistence of AF and SC is also found in CeRh$_{1-x}$Co$_x$In$_5$ \cite{Jeffries}.
In Ce$_{1-y}$La$_y$RhIn$_5$ the magnetic order vanishes for $y >$ 0.4 \cite{Pagliuso2}. Beyond that doping level it remains a paramagnet where short-range magnetic correlations are observed \cite{Light,Correa}.

However, doping and pressure are not the only tuning parameters. A magnetic field $B$, for instance, suppresses superconductivity resulting in a metallic ground state when the kinetic enegy of the induced screening currents exceeds the SC condensation energy. This orbital limit is characterized by a first order phase transition at the critical field $B_c$ in type-I superconductors, or by a second order transition at the upper critical field $B_{c2}$ when the magnetic pressure is continuously relaxed through the vortex mixed state of type-II superconductors.
Other phenomena caused by an applied field can also suppress superconductivity. In the Pauli or paramagnetic limit, the singlet state formed by the pairs is polarized by an external $B$ when the Zeeman energy of the of the partially polarized spins overcomes the condensation energy, breaking the pairs and destroying SC.
Whether a superconductor is orbital or Pauli limited can be characterized by the so called Maki parameter, $\alpha = \sqrt{2} B_o / B_p$ \cite{Maki1}, where $B_o$ and $B_p$ are the orbital and paramagnetic critical fields, respectively. 

Two striking predictions were made in the pure paramagnetic limit ($\alpha \rightarrow \infty$). First, the phase transition at $B_{c2}$ should change from second to first order below a critical temperature $T_0$ \cite{Maki2}. Second, Fulde and Ferrel \cite{FF} and Larkin and Ovchinnikov \cite{LO} proposed a new inhomogeneous superconducting state (FFLO state) in which the superconducting order parameter is modulated along the magnetic field direction  developing nodes where normal electrons take advantage of the Zeeman energy and become polarized. 
Even when orbital effects are present the FFLO state can be realized for $T < T_0$ and close to $B_{c2}$, as long as the paramagnetic effect is dominant ( $\alpha >$ 1.8) \cite{GG}.

The large $\alpha$ required for the formation of the FFLO state can be achieved either by a high $B_o$ and/or a low $B_p$. Systems with heavy quasiparticle mass which reduces the kinetic energy of the shielding currents or with  two-dimensional character which reduces the electrons orbital degrees of freedom will push $B_o$ up. A high Pauli susceptibility is also indicative of the required paramagnetic character.

CeCoIn$_5$ meets all these requirements and has in addition a large mean free path placing the system in the clean limit \cite{Movshovich}. In fact, a crossover from a second to first order transition was observed in $B_{c2}$ below $T_0 \sim$ 1 K suggesting Pauli limited SC \cite{Murphy,Tayama,Bianchi1}. Finally, specific heat experiments have recently shown \cite{Radovan} a second order-like transition that occurs at low temperature ($T \lesssim$ 0.35 K) just below $B_{c2}$ that is believed to be the first ever clear observation of the FFLO phase. Different experiments have confirmed this observation \cite{Bianchi2,Watanabe,Capan,Martin,Kakuyanagi,Miclea}.
A good agreement between the different works is found in the overall magnetic field versus temperature phase diagram, see Fig. \ref{fig3}(a). However, as with any proposed new state of matter, some debate and several open questions about the character and nature of it arise, showing the need for new experimental information pending a conclusive proof of the order parameter by means of microscopic techniques capable of direct observation of the proposed spatially inhomogeneous superconducting state.

In this work we study the coupling of the order parameter to the lattice through magnetostriction experiments. 
Besides its amazing sensitivity, linear magnetostriction (as well as thermal-expansion) is a powerful technique to study anisotropy and reduced dimensionality because each crystal axis can be measured independently for different directions of the applied magnetic field $B$. 
Our results show that the lattice coupling is strongly anisotropic and confirm the two-dimensional character of this phase. These observations in conjunction with previous results are consistent with a FFLO state and inconsistent with field induced magnetic order.

\begin{figure}[b]
  \includegraphics[width=\columnwidth]{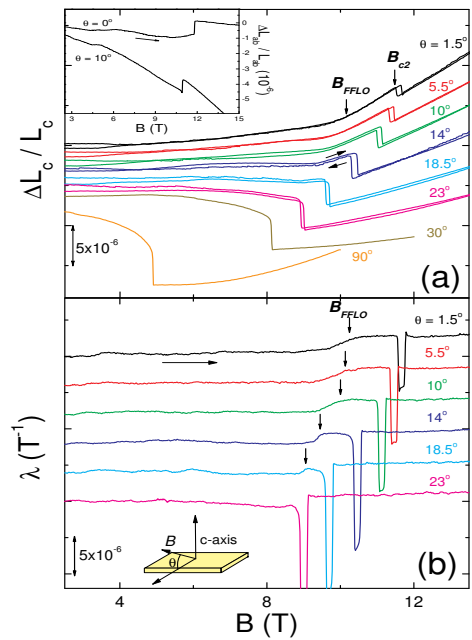}
  \caption[]{(a) Linear c-axis magnetostriction versus field at $T \approx$ 30 mK for different directions of the applied magnetic field. Inset: ab-plane linear magnetostriction. (b) c-axis magnetostriction coefficient $\lambda = \frac {1} {L} ( \frac {\partial L} {\partial B})$. Inset: sketch of the experimental configuration. Curves have been  vertically shifted.}
  \label{fig1}
\end{figure}

CeCoIn$_5$ single crystals were grown by the self-flux technique. The iso-thermal linear magnetostriction experiments were performed on a 1$\times$1$\times$1.5 mm$^3$ sample using a titanium capacitance dilatometer \cite{Schmiedeshoff} with a resolution $\sim$ 0.3 \AA \, ($\Delta L / L \sim$ 10$^{-8}$). The dilatometer is placed inside the mixing chamber of a dilution refrigerator immersed directly in the He$^3$-He$^4$ mixture, achieving an excellent iso-thermal condition ($\pm$ 2 mK) and a base temperature close to 25 mK. The results were verified with a second larger sample.
A small asymmetry is observed in the slope of the magnetostriction curves with respect to the ab-planes ($\theta$ = 0$^o$ in our experimental configuration sketched in Fig. \ref{fig1}(b)), i.e. between negative and positive angles, whose origin is not yet understood \cite{torque}. Our results and conclusions, however, are independent of this artifact.

Figure \ref{fig1}(a) displays the c-axis linear magnetostriction $(\frac {L(B) - L(0)}  {L(0)})$ for different directions of $B$ at $T \approx$ 30 mK.
Two features are clearly distinguishable. A higher field first order transition which monotonically moves to lower $B$ as the field is rotated towards the c-axis ($\theta$ = 90$^o$). This discontinuity occurs at $B$ = 11.7 T for $\theta$ = 0$^o$ and $B$ = 4.9 T for $\theta$ = 90$^o$ confirming that it corresponds to the upper critical field $B_{c2}$ \cite{Murphy}. The length change at $B_{c2}$ also increases continously as the field moves away from the ab-planes reaching a value of 4 $\times$ 10$^{-6}$ ($\theta$ = 90$^o$) in good agreement with previous results \cite{Bianchi1,Takeuchi}. 
At lower fields a second feature appears. It is a second order-like anomaly that occurs at $B \sim$ 10 T for $\theta \rightarrow$ 0 and is observed only at low angles ($\theta \lesssim$ 20$^o$). This can be clearly seen in Fig. \ref{fig1}(b) where the field dependence of the c-axis magnetostriction coefficient  $\lambda = \frac {1} {L} ( \frac {\partial L} {\partial B})$ is shown. $B_{c2}$ and the second order transition are detected as a peak and a "jump", respectively. 
The hysteresis between the up-sweep and down-sweep curves is appreciable at $B_{c2}$ but negligible at the lower field transition.

\begin{figure}[t]
  \includegraphics[width=\columnwidth]{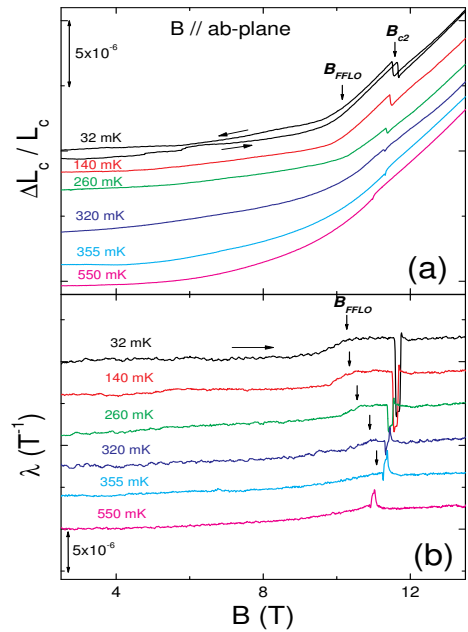}
  \caption[]{(a) Linear c-axis magnetostriction versus field at different temperatures. The magnetic field lies along the ab-planes. (b) c-axis magnetostriction coefficient. Curves have been vertically shifted.}
  \label{fig2}
\end{figure}

\begin{figure}[t]
  \includegraphics[width=\columnwidth]{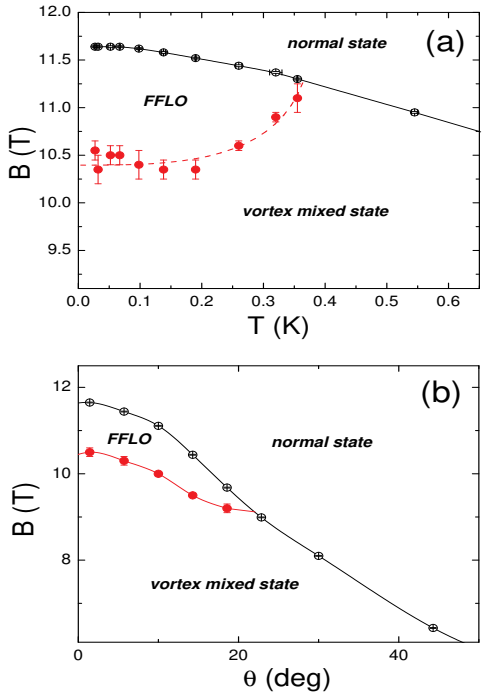}
  \caption[]{(a) Magnetic field versus temperature ($B \parallel$ ab-planes) and (b) magnetic field versus angle ($T \approx$ 30 mK) phase diagrams. The lines are guides to the eye.}
  \label{fig3}
\end{figure}

The temperature evolution of both transitions can be observed in Fig. \ref{fig2}. The upper panel shows the c-axis magnetostriction for in-plane fields ($\theta$ = 0$^o$) at different temperatures, while the lower panel shows the magnetostriction coefficient $\lambda$. The second order transition (arrows in Fig. \ref{fig2}(b)) moves to higher fields as the temperature is increased and vanishes around 0.35 K.  Above this temperature only a peak associated with $B_{c2}$ is observed. This peak as well as the hysteresis become smaller as the temperature is raised implying an evolution to a conventional second order critical field as has already been reported \cite{Murphy,Tayama,Bianchi1}.

Our results are summarized in the phase diagrams shown in Fig. \ref{fig3}. The in-plane $B - T$ phase diagram displayed in Fig. \ref{fig3}(a) is in very good agreement with previous works, including the area of occurrence of the proposed FFLO state \cite{Radovan,Bianchi2,Watanabe,Capan,Martin} and the almost linear $T$-dependence of $B_{c2}$ down to very low temperatures. This nearly linear T-dependence is interpreted as a magnetically enhanced SC due to the predominantly paramagnetic character \cite{Tayama,Radovan,Won}. In this scenario, our lower field second order transition corresponds to the transition from the vortex mixed state to the FFLO state, $B_{FFLO}$.
Fig. \ref{fig3}(b) shows the $B - \theta$ phase diagram at $T \approx$ 30 mK. The reduced angular range ($\theta \lesssim$ 20$^o$) where the proposed FFLO state is observed confirms the quasi two-dimensional character of this phase and was attributed to the planar crystal structure that partially inhibits the orbital motion along the c-axis \cite{Radovan,Martin}. 
The results in this paper provide further evidence of reduced dimensionality and anisotropy. The inset of Fig. \ref{fig1}(a) shows the ab-plane linear magnetostriction for fields close to the parallel configuration ($\theta$ = 0 where $B \parallel L_{ab}$). No second order anomaly is detected below the sharp transition associated with $B_{c2}$. 
Within the original prediction of the FFLO state \cite{FF,LO} consisting in planes of normal electrons perpendicular to the field, our results state that a lattice coupling exists only along these nodal planes but not perpendicular to them \cite{comment}.

Finally, we compare our results with the predictions of an alternative explanation for the proposed FFLO state that considers this field induced phase as a magnetically ordered state \cite{Flouquet}.
The argument for this state is based on the pressure versus temperature phase diagram of CeRhIn$_5$ \cite{Fisher,Mito,Llobet,Nicklas,Knebel} and is as follows. The AF ordering temperature is continously depressed with pressure and coexists with SC for $P \gtrsim$ 10 kbar. Beyond $P_{c1} \approx$ 18 kbar, where $T_N(P_{c1}) = T_c(P_{c1})$, AF abruptly disappears and only SC is observed. Thus, the experimental evidence shows that SC and AF coexist in CeRhIn$_5$ as long as $T_N(P) \geq T_c(P)$. 
However, a magnetic field should weaken SC moving the SC+AF/SC boundary to higher pressures, thereby enhancing the relative strength of the magnetic correlations and probably inducing a magnetically ordered phase even when $T_N(P,B) < T_c(P,B)$.
Very recently, this prediction has been confirmed \cite{Park}. A field induced, magnetically ordered phase whose $B -T$ phase diagram resembles the one in Fig. \ref{fig3}(a) is observed between $P_{c1}$ and $P_{c2}$ = 22.5 kbar. $P_{c2}$ characterizes a quantum critical point (QCP) at which $T_N(P,B)$ goes to zero and no magnetic order is observed above it.
CeCoIn$_5$ is known to be in the vicinity of an AF QCP \cite{Nicklas,Sishido,Sidorov,Paglione,Bianchi3}. Thus, if CeCoIn$_5$ at ambient pressure is located between $P_{c1}$ and $P_{c2}$, the proposed FFLO state may correspond instead to a field induced magnetic order.  

The experimental results are not consistent with this interpretation. First, at high enough fields $T_N(B)$ exceeds $T_c(B)$. That means that in the phase diagram the field induced transition should cross the upper critical field, as is observed in CeRhIn$_5$ \cite{Park}. Our high sensitivity experiment reveals no signature of $B_{FFLO}$ above $B_{c2}$, as is also observed in experiments performed sweeping $T$ at constant $B$ \cite{Bianchi2}.
Second, the area occupied by this magnetically ordered phase should decrease with $P$ as is also reported in CeRhIn$_5$ \cite{Park}. However, recent high pressure specific heat experiments in CeCoIn$_5$ show that this area increases with $P$. Third, magnetic ordering should be, in principle, observed for any field direction and it is not.
These reasons lead us to conclude that the low temperature high field phase of CeCoIn$_5$ is most consistent with the FFLO state and not field induced magnetic order.   
 
We gratefully acknowledge K. Yang for a careful reading of the manuscript, M. Case for helpful discussions, and D. McIntosh and R. Newsome for technical assistance. Work at the NHMFL was performed under the auspices of the National Science Foundation, the State of Florida and NNSA DE-FG52-03NA00066. 
 
\section{references}


\begin{thebibliography}{99}
\bibitem{torque}  The same asymmetry is observed in the second sample 
whose mass is 15 times larger than the first one.
On the other hand, there is a perfect scaling with the sample length for 
both samples. Hence, we can rule out a spurious dilation due to torque on 
the sample arising from a misalignment between magnetic moment and 
field that would mean a volume/mass scaling.
\bibitem{comment} In fact, our results prove a FFLO-lattice coupling only 
along the c-axis. But according to the pressure studies in Ref. \cite{Miclea} 
and from thermodynamics considerations we can predict an ab-plane ($L_{ab}$)
lattice effect when the magnetic field lies also along the ab-planes but 
perpendicular to $L_{ab}$.
\end{thebibliography}
\end{document}